\documentclass[]{spie}  %>>> use for US letter paper
\usepackage{amsmath,amsfonts,amssymb}
\usepackage{graphicx}
\usepackage[table,xcdraw]{xcolor}
\usepackage{siunitx}
\usepackage[colorlinks=true, allcolors=blue]{hyperref}

\title{The Space Coronagraph Optical Bench (SCoOB): 8. end-to-end numerical modeling of the testbed to estimate the contrast limits}

\author[a]{Ramya M Anche}
\author[a]{Kyle Van Gorkom}
\author[a,c]{Kian Milani}
\author[a,c]{Kevin Derby}
\author[a,c]{Emory Jenkins}
\author[b]{Jaren Ashcraft}
\author[a]{Saraswathi Kalyani Subramanian}
\author[a]{Patrick Ingraham}
\author[c]{Daewook Kim}
\author[c]{Heejoo Choi}
\author[a]{Olivier Durney}
\author[a]{Ewan Douglas}

\affil[a]{Steward Observatory, University of Arizona, 933N Cherry Avenue, Tucson, Arizona, 85721, USA}
\affil[b]{Department of Physics, University of California, Santa Barbara, CA 93106}
\affil[c]{James C. Wyant College of Optical Sciences, University of Arizona, 933N Cherry Avenue, Tucson, Arizona, 85721, USA}

\authorinfo{}

\begin{document} 
\maketitle

\begin{abstract}
The space coronagraph optical bench (SCoOB) at the University of Arizona is a high-contrast imaging testbed designed to operate in a vacuum to obtain a contrast better than $10^{-8}$ in optical wavelengths using vector vortex coronagraph (VVC) masks. The testbed performance in a half-sided D-shaped dark hole is $2.2\times10^{-9}$ in a $\ll 1 \%$ BW, $4\times10^{-9}$ in a 2\% BW, and $2.5\times10^{-8}$ in a 15\% BW. While the testbed has met the design specification contrast requirements in monochromatic wavelengths, comprehensive end-to-end numerical modeling to assess contrast limits across different bandpasses has yet to be conducted. In this work, we discuss the results of numerical modeling for the SCoOB testbed in both monochromatic and 10\% bandwidths at 525 nm and 630 nm. This modeling incorporates measured VVC retardance, modeled polarization aberrations, measured surface and reflectivity errors, and diffuse and surface reflectivity. We explore and discuss the various factors contributing to the contrast limits.
\end{abstract}

% Include a list of keywords after the abstract 

\section{INTRODUCTION}
\label{sec:intro}  % \label{} allows reference to this section
High-contrast imaging testbeds serve as experimental platforms to test various coronagraphs and wavefront sensing and control algorithms that are intended to be used in the high-contrast imaging instruments for future space-based telescopes aiming to directly image Earth-like planets around Sun-like stars. The Space Coronagraph Optical Bench (SCoOB) testbed \cite{van2022space,ashcraft2022space,gorkom_scoob_2024} is a high-contrast imaging testbed operational since 2021, achieving a contrast of $2.2\times10^{-9}$ in a $\ll 1 \%$ BW and $2.5\times10^{-8}$ in a 15\% BW in a $3-10\lambda/D$ dark hole (DH) at 630nm using a VVC \cite{gorkom_scoob_2024}. SCoOB employs implicit Electric Field Conjugation \cite{iefc} as the High-order Wavefront sensing and Control (HOWFSC) algorithm to perform DH digging through the use of pairwise probes on a deformable mirror (DM) with Hadamard modes. 
Additionally, another HOWFSC technique, known as the self-coherent camera, has recently been demonstrated on SCoOB, achieving a contrast of $10^{-8}$ in monochromatic wavelength of 630nm \cite{derby2025scoob}. The Lyot-based low-order wavefront sensing and control (LOWFSC) in combination with HOWFSC has also been demonstrated on SCoOB, maintaining a contrast of $10^{-8}$ contrast levels in air \cite{milani2025scoob}, effectively correcting the low-order aberrations. 

To understand the various factors contributing to the contrast limitation of SCoOB, previous modeling efforts have been carried out by Anche et al \cite{anche_pol_scoob_2024} and Ashcraft et al \cite{ashcraft_mueller_scoob_2024} to estimate the polarization aberrations and their effect on the contrast performance. Additionally, Van Gorkom et al \cite{van2025performance} assessed the contrast limits of SCoOB in UV wavelengths using end-to-end numerical modeling. In this paper, we present the contrast budget for SCoOB in optical wavelengths and calculate the contrast limits in 2\%, 5\%, and 10\% bandpasses centered at 630nm and 543 nm. Section \ref{sec:optical_layout} shows the optical layout of SCoOB, highlighting its key components. An overview of the simulations is provided in section \ref{sec:overview_sims}, and the contrast budget is explained in detail in section \ref{sec:contrast_budget}. Finally, we summarize the results in section \ref{sec:conclusions}.
\section{SCoOB Optical layout and contrast performance}
\begin{figure}[!ht]
    \centering
    \includegraphics[width=1\linewidth]{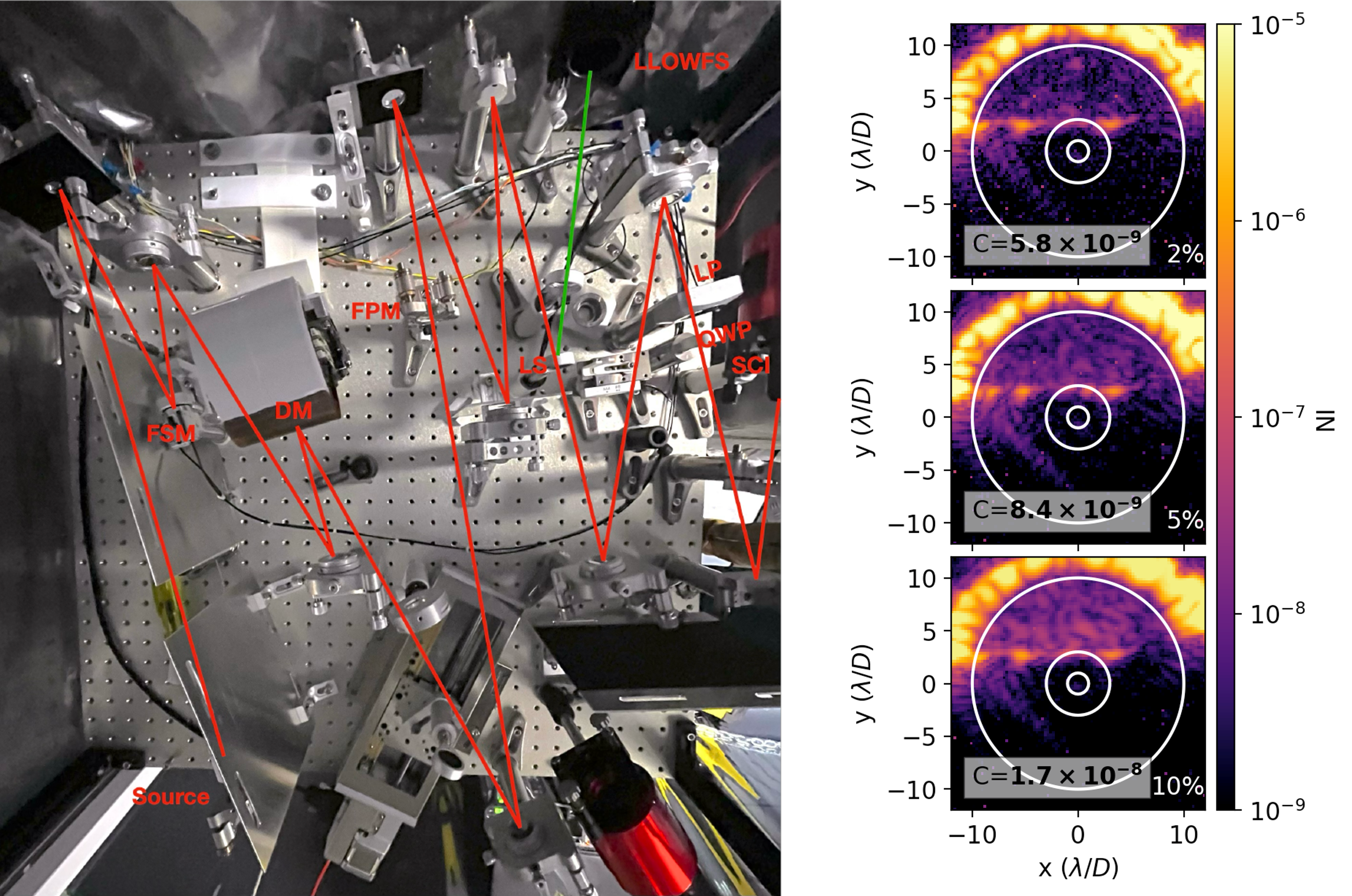}
    \caption{Left: Optical layout of SCoOB showing the source, FSM, OAPs, DM, FPM, LOWFSC arm and the science camera. Right: The best contrast performance of SCoOB is $2.2\times10^{-9}$ contrast in quasi-monochromatic light, $5.8\times10^{-9}$ in a 2\% bandwidth, $8.4\times10^{-9}$ in a 5\% bandwidth, and $1.7\times10^{-8}$ in a 10\% bandwidth centered at 630nm in a D-shaped, half-sided DH from $3-10\lambda/D$.}
    \label{fig:opt_layout}
\end{figure}

The optical layout of SCoOB on the breadboard is shown in Figure \ref{fig:opt_layout}. The light from the fiber is collimated and then refocused on the micron-fabricated pinhole of 5$\mu$m. The input linear polarizer (LP) and the quarter-wave plate (QWP) retarder are placed in the collimated beam before the pinhole (not shown in Figure \ref{fig:opt_layout}). The first off-axis parabola (OAP) collimates the beam and reimages the pupil onto the Fast Steering Mirror (tip/tilt mirror), which is reimaged onto the deformable mirror plane using another pair of OAPs. The vector vortex waveplate is placed at the first focal plane downstream of the DM. A pair of fold mirrors and an OAP create the pupil at the Lyot stop (LS) location, where the ring of fire outside the LS is reflected to the LLOWFSC arm, and the main science beam is focused onto the science camera. The output LP and QWP are placed in the collimated beam just before the science camera.
\label{sec:optical_layout}
\section{Overview of end-to-end simulations}
\label{sec:overview_sims}
To establish the performance limits of the testbed, we developed an end-to-end numerical diffraction model\cite{van2025performance} in \href{https://hcipy.org/}{hcipy}. This model incorporates surface and reflectivity errors on all optics, quantization error and actuator noise on the DM, fabrication errors on the VVC, polarization aberrations from all the optics and specular and diffuse reflections from stops at the beginning of the optical train. The model is run for two central wavelengths in optical region at 630nm and 543nm to estimate the contrast performance and contrast budget in both monochromatic and 2-10\% bandwidths.  
\section{Contrast budget}
\label{sec:contrast_budget}
\begin{figure}[!ht]
    \centering
    \includegraphics[width=1\linewidth]{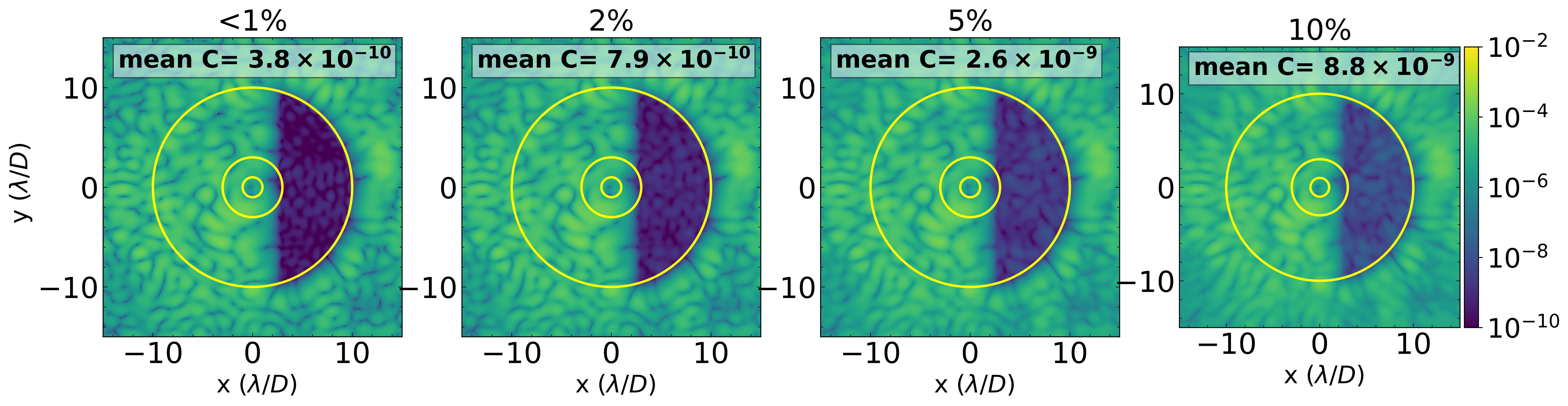}
    \caption{The contrast vs the focal plane distance is shown for the EFC chromatic residuals for central wavelength of 630nm with 2-10\% bandwidths incorporating surface and reflectivity errors from all the surfaces.}
    \label{fig:efc_630nm}
\end{figure}

\begin{figure}[!ht]
    \centering
    \includegraphics[width=1\linewidth]{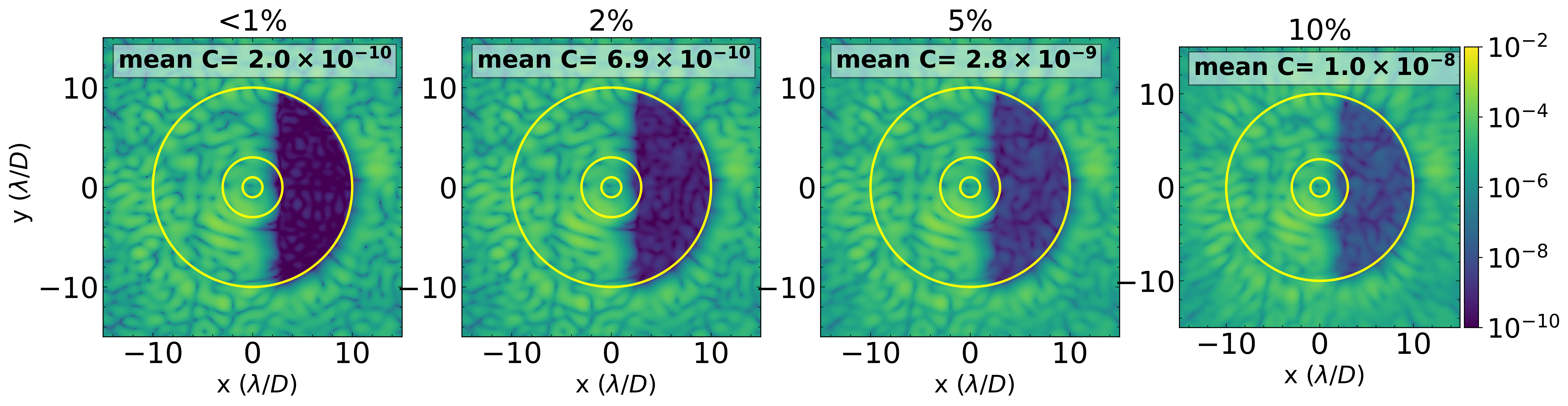}
    \caption{The contrast vs the focal plane distance is shown for the EFC chromatic residuals for central wavelength of 543nm with 2-10\% bandwidths incorporating surface and reflectivity errors from all the surfaces.}
    \label{fig:efc_543nm}
\end{figure}

\textbf{Chromatic EFC Residuals:} Surface errors are generated from realizations of power spectral densities (PSDs), which have been fit to surface measurements of OAPs and flats with a 4D PhaseCam interferometer and a white-light coherence scanning microscope. Reflectivity errors have not been directly measured but are approximated using PSD parameters used for other high contrast imaging instruments\cite{mendillo_tolerances}. The contrast floor set by surface errors arises primarily from the Talbot effect, which causes aberrations in propagated electric fields to cycle between phase and amplitude at characteristic lengths that depend on both spatial frequency and wavelength\cite{Talbot_1836,Shaklan06, pueyo_polychromatic_2007, mazoyer_pueyo}. The EFC chromatic residuals are estimated for different bandwidths as shown in Figures \ref{fig:efc_630nm} and \ref{fig:efc_543nm}.

\textbf{DM quantization and actuator noise:}  DM quantization error arises from the finite bit depth of the DM electronics, which sets a minimum actuator step size\cite{ruane_2021_quantization}. DM actuator noise is a similar term that arises from voltage noise of the electronics, assumed to follow a Gaussian distribution\cite{van2025performance}. We adopt values for a Kilo-C DM with 16-bit electronics, 1 LSB RMS voltage noise, and operating at a 40\% voltage bias to reduce the actuator gain while retaining enough stroke to correct static instrument errors and perform electric field conjugation (EFC)\cite{efc}. The contrast floor set by these terms is estimated both analytically and by incoherently summing many realizations of quantization error and actuator noise.

\begin{figure}[!ht]
    \centering
    \includegraphics[width=1\linewidth]{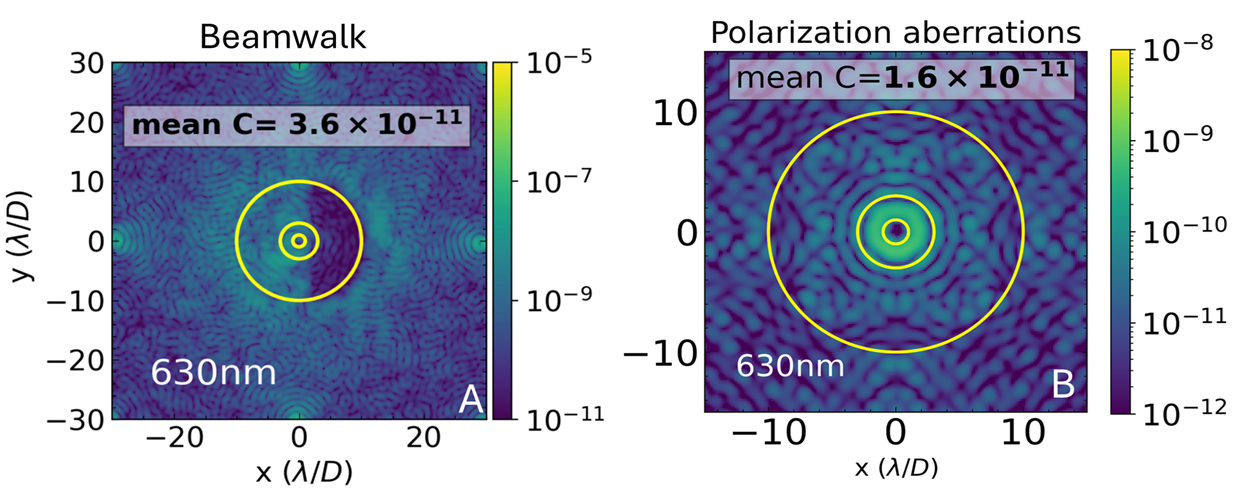}
    \caption{A:The contrast vs the focal plane distance is shown for SCoOB in the presence of $3\times10^{-3} \lambda/D$ jitter at 630nm. B. The contrast vs the focal plane distance is shown for SCoOB at 630nm in the presence of polarization aberrations.}
    \label{fig:beamwalk_polab}
\end{figure}

\textbf{Jitter and Beamwalk:} Beamwalk is a beam shear on optical surfaces that arises from line of sight jitter, the net effect of which is a dynamic high spatial-frequency error that is assumed to uncorrected by the high order control loop. SCoOB jitter in vacuum was previously measured at $3\times10^{-3} \lambda/D$ RMS at 630nm. We simulated the contrast floor from this term by digging a DH in the absence of jitter and then injecting many realizations of jitter pulled from a Gaussian distribution at the first optic in the testbed. The intensity at the focal plane is incoherently summed to estimate the beamwalk-limited contrast. Figure \ref{fig:beamwalk_polab}A shows the beamwalk limited contrast estimated for SCoOB at 630nm.

\textbf{Polarization aberrations:} Polarization aberrations originate from the difference in the Fresnel reflection coefficients from all the optical surfaces \cite{anche_2023,Breckinridge_2015,ashcraft2025}. We model the polarization aberrations by performing a polarization ray trace to estimate the Jones pupils at the exit pupil. We propagate the Jones pupils thorough the VVC to estimate the polarization-aberration limited contrast. End-to-end simulations of polarization aberration of SCoOB is presented in detail by Anche et al \cite{anche_pol_scoob_2024} and the polarization aberration limited contrast is shown in Figure \ref{fig:beamwalk_polab}B at 630nm. 

\begin{figure}[!ht]
    \centering
    \includegraphics[width=0.495\linewidth]{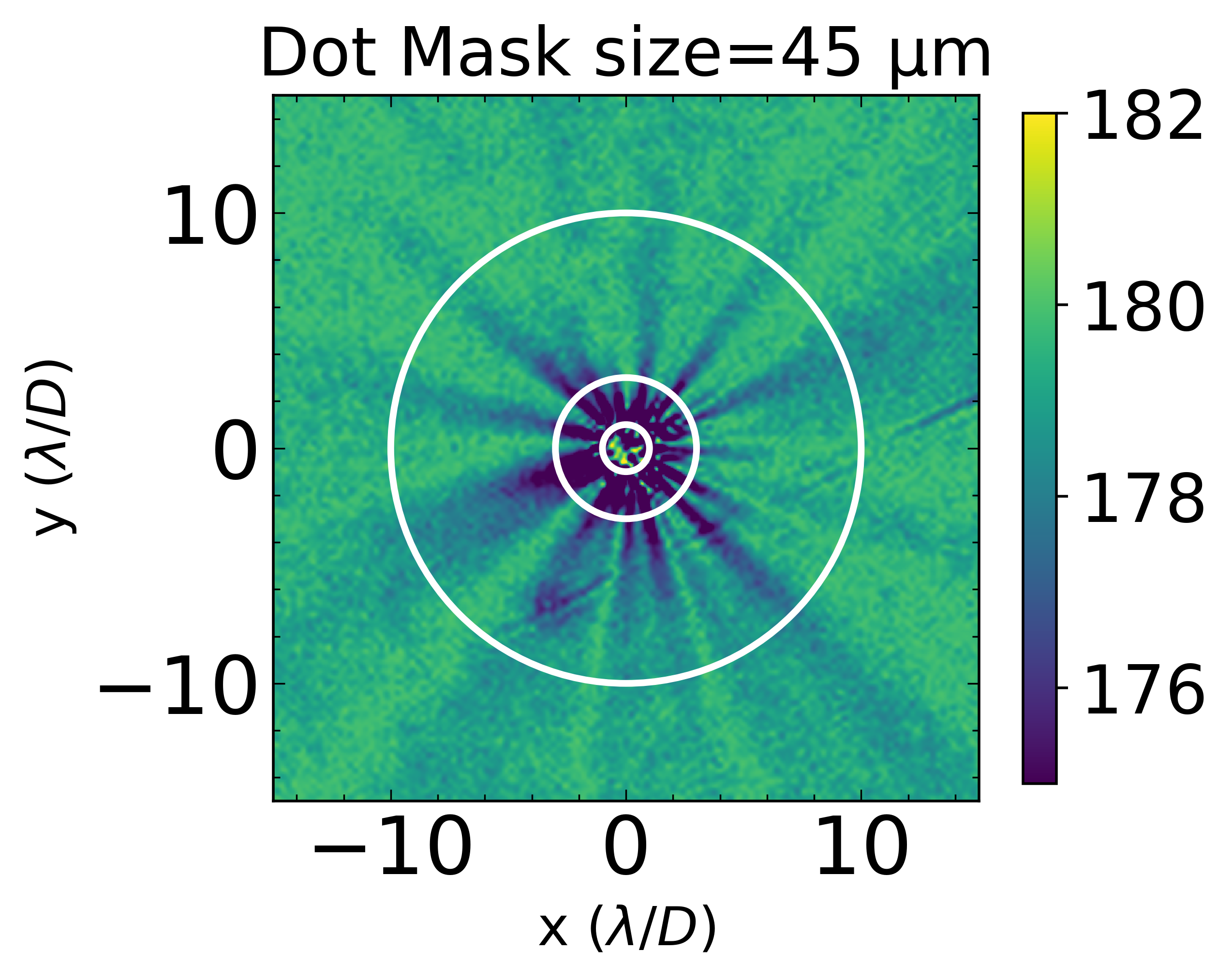}
    \includegraphics[width=0.495\linewidth]{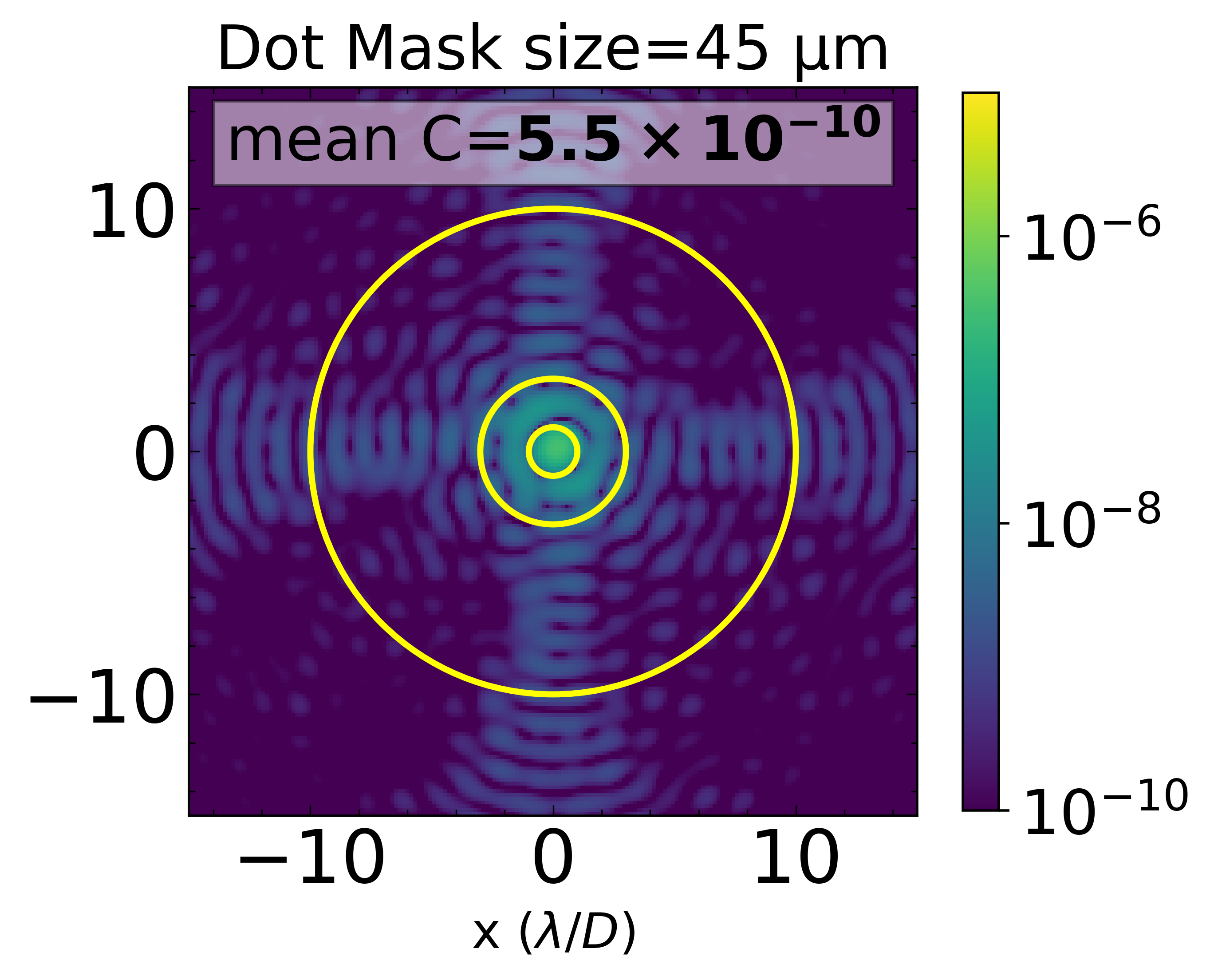}
    \caption{Left: The measured retardance map of one of the Beamco VVCs showing the spatially variation of retardance. Right: The contrast vs the focal plane distance estimated for the Beamco VVC with a dot mask of 45 $\mu$m.}
    \label{fig:vvc_fab_errors}
\end{figure}

\textbf{VVC mask fabrication errors:} As a VVC is an achromatic half-wave retarder, it is expected to have a retardance of 180\textdegree. However, due to manufacturing errors, the retardance on the VVC varies spatially as shown in the Figure \ref{fig:vvc_fab_errors}. We estimate the contrast performance with the measured retardance from a VVC manufactured by Beam Co to incorporate the retardance errors shown in Figure \ref{fig:vvc_fab_errors}. 

\textbf{Specular and diffuse reflectivity:} The contrast floor due to the finite reflectivity of the ``non-reflective'' portions of the pupil stops is computed following the approach adopted for Roman CGI\cite{balaMilestone1,Balasubramanian_2015,Balasubramanian_2019}. The specular component of the mask reflection is propagated through the diffraction model to estimate the focal-plane contribution. The diffuse component is calculated by treating it as a Lambertian scatterer and estimating the fraction of the diffuse reflection that is scattered to small angles (corresponding to the dark hole region). The reflection from two surfaces is included in this analysis: an Acktar metal velvet stop at OAP0 with an estimated $10^{-6}$ specular reflectance and $1\%$ diffuse reflectance, and an Aeroglaze Z306 stop at the FSM with an estimated $10^{-6}$ specular component and $4.8\%$ diffuse component. The terms in the contrast budget are summarized in Table \ref{tab:contrast_budget} at two central wavelengths and in bandwidths from 2-10\%.
\section{Conclusions and Discussions}
\label{sec:conclusions}
The predicted contrast floor is dominated by diffuse reflection from the Aeroglaze Z306 stop at the FSM in narrow bandwidths ($\leq5\%$) and by chromatic EFC residuals in broader bandwidths ($\geq10\%)$. We plan to mitigate the narrowband limit by replacing the Aeroglaze Z306 stop with either a black silicon (BSi) or carbon nanotube (CNT) stop, both of which show $<1\%$ hemispherical reflectance in visible wavelengths\cite{butler_2010,georgiev_2019}. The contrast limit due to chromatic EFC residuals can only be mitigated with a significant overhaul of the testbed design or improved surface quality on optical components, although pairwise probing for EFC at multiple sub-bands may be able to improve broadband performance\cite{zhou_efc_roman_2016}. 
\begin{table}[!ht]
\begin{center}
\begin{tabular}{|l|l|l|}
\hline
\rowcolor[HTML]{19194D} 
{\color[HTML]{FFFFFF} \textbf{Terms}}                     & {\color[HTML]{FFFFFF} \textbf{630nm}}                                                                                                      & {\color[HTML]{FFFFFF} \textbf{543nm}}                                                                                                    \\ \hline
\rowcolor[HTML]{E2EBFD} 
Chromatic EFC residuals (surface and reflectivity errors) & \begin{tabular}[c]{@{}l@{}}2\% : 7.9X$10^{-10}$\\ 5\% : 2.6X$10^{-9}$\\ 10\%: 8.8X$10^{-9}$\end{tabular}                                   & \begin{tabular}[c]{@{}l@{}}2\% : 6.9X$10^{-10}$\\ 5\% : 2.8X$10^{-9}$\\ 10\%: 1.0X$10^{-8}$\end{tabular}                                 \\ \hline
\rowcolor[HTML]{E2EBFD} 
Jitter and beamwalk                                       & 3.6X$10^{-11}$                                                                                                                             & 2.95X$10^{-11}$                                                                                                                          \\ \hline
\rowcolor[HTML]{E2EBFD} 
Polarization aberrations                                  & 1.6X$10^{-11}$                                                                                                                             & 1.6X$10^{-11}$                                                                                                                           \\ \hline
\rowcolor[HTML]{E2EBFD} 
VVC fabrication errors                                    & 5.5X$10^{-10}$                                                                                                                             & 1.0X$10^{-9}$                                                                                                                            \\ \hline
\rowcolor[HTML]{B4D5E5} 
DM quantization                                           & 1.6X$10^{-11}$                                                                                                                             & 2.1X$10^{-11}$                                                                                                                           \\ \hline
\rowcolor[HTML]{B4D5E5} 
DM noise                                                  & 1.8X$10^{-10}$                                                                                                                             & 2.4X$10^{-10}$                                                                                                                           \\ \hline
\rowcolor[HTML]{B4D5E5} 
Specular + diffuse reflectivity                           & 2.9X$10^{-9}$                                                                                                                              & 2.2X$10^{-9}$                                                                                                                            \\ \hline
\rowcolor[HTML]{19194D} 
{\color[HTML]{FFFFFF} \textbf{Total}}                     & {\color[HTML]{FFFFFF} \textbf{\begin{tabular}[c]{@{}l@{}}2\% : 4.4X$10^{-9}$\\ 5\% : 6.2X$10^{-9}$   \\ 10\%: 1.2X$10^{-8}$\end{tabular}}} & {\color[HTML]{FFFFFF} \textbf{\begin{tabular}[c]{@{}l@{}}2\% : 4.1X$10^{-9}$\\ 5\% : 6.4X$10^{-9}$\\ 10\%: 1.36X$10^{-8}$\end{tabular}}} \\ \hline
\end{tabular}
\caption{Contrast budget for SCoOB central wavelengths $\lambda$ from \si{543} and \SI{630}{\nano \meter}, and bandwidths from 2-10\%. The dark hole is a D-shaped, half-sided DH from $3-10\lambda/D$. The individual terms are incoherently added to produce the total contrast floor.}
\label{tab:contrast_budget}
\end{center}
\end{table}

\appendix    %>>>> this command starts appendixes

\acknowledgments % equivalent to \section*{ACKNOWLEDGMENTS}       
Portions of this research were supported by funding from the Technology Research Initiative Fund (TRIF) of the
Arizona Board of Regents and by generous philanthropic donations to the Steward Observatory of the College of Science at the University of Arizona.

% References
\bibliography{report, report_uv} % bibliography data in report.bib
\bibliographystyle{spiebib} % makes bibtex use spiebib.bst

\end{document}